\documentclass[12pt]{article}
\usepackage{epsf}
\usepackage[cp1251]{inputenc}

\begin{document}

\title{
{\bf Nonlinearity of vacuum Reggeons and exclusive diffractive 
production of vector mesons at HERA}}
\author{{\it A.A. Godizov\thanks{godizov@sirius.ihep.su}, 
V.A. Petrov}\thanks{Vladimir.Petrov@ihep.ru}\\
{\small {\it Institute for High Energy Physics, 142281 Protvino, Russia}}}
\date{}
\maketitle

\vskip-1.0cm

\begin{abstract}
The processes of exclusive photo- and electroproduction of vector 
mesons $\rho^0$(770), $\phi$(1020) and $J/\psi$(3096) at collision 
energies $30\,GeV<W<300\,GeV$ and transferred momenta squared 
$0<-t<2\,GeV^2$ are considered in the framework of a 
phenomenological 
Regge-eikonal scheme with nonlinear Regge 
trajectories in which their QCD asymptotic behavior is taken 
into account explicitly. By comparison of available experimental 
data from ZEUS and H1 Collaborations with the model predictions 
it is demonstrated that corresponding angular distributions 
and integrated cross sections in the above-mentioned kinematical 
range can be quantitatively described with use of two $C$-even vacuum 
Regge trajectories. These are the 
``soft'' Pomeron dominating the high-energy reactions without a 
hard scale and the ``hard'' Pomeron giving an essential 
contribution to photo- and electroproduction of heavy vector mesons 
and deeply virtual electroproduction of light vector mesons. 
\end{abstract}

\vspace*{1cm}

\section*{Introduction}

The aim of this paper is to demonstrate that reactions of exclusive 
vector meson photo- and electroproduction can be described in 
the framework of a Regge-eikonal scheme 
(the general Regge-eikonal approach was developed in \cite{arnold}) 
similar to that as for high-energy nucleon-nucleon scattering. 
One of the distinctive features of the proposed model is the use 
of nonlinear parametrizations for Regge trajectories in the 
diffraction region in which their QCD asymptotic 
behavior is taken into account explicitly. 
In essence, the behavior of Regge trajectories at large transferred 
momenta is a fact of fundamental 
importance: QCD does not permit linear trajectories. Hence, 
there emerges the problem of compatibility of nonlinearity of 
Regge trajectories following from QCD with available set of 
experimental data. The scheme proposed in \cite{godizov} 
and applied in this paper partially solves this problem 
since in the framework of our approach phenomenological 
Regge trajectories in the diffraction region satisfy 
the QCD asymptotic relations explicitly. 

The detailed discussion 
of the fundamental properties of Regge trajectories (such as their 
essential nonlinearity in the Euclidean domain) 
can be found in \cite{godizov}. Here we limit ourselves by substantiating 
several useful relations. 

Our viewpoint (by no means very original) is based on the 
conviction that QCD is the fundamental theory of strong interaction. 
So, in the limit of large negative values of the argument 
(equivalent to short distances) the exchange 
by any Reggeon must turn into exchange by some colour-singlet 
parton combination and the asymptotic behavior of Regge poles 
corresponding to exchanges by definite parton combinations 
can be determined. 
For example, if some Reggeon corresponds to 
exchange by the quark-antiquark pair ($f_2$ Reggeon, etc.) 
in the range of the perturbative QCD validity we come to 
\cite{kwiecinski}
\begin{equation}
\label{meson}
\alpha_{\bar qq}(t) = 
\sqrt{\frac{8}{3\pi}\alpha_s(\sqrt{-t})}+o(\alpha_s^{1/2}(\sqrt{-t}))
\end{equation}
In the case of multigluon exchange 
one obtains \cite{wu} 
\begin{equation}
\label{gluon}
\lim_{t\to -\infty}\alpha_{gg...g}(t) = 1
\end{equation}
and, in particular, for two-gluon exchange \cite{kirschner} 
\begin{equation}
\label{hard}
\alpha_{gg}(t) = 1+\frac{12\,\ln 2}{\pi}\alpha_s(\sqrt{-t})
+o(\alpha_s(\sqrt{-t}))
\end{equation}
where $\alpha_s(\mu)\equiv g_s^2(\mu)/4\pi$ is the QCD running 
coupling.

Besides, if we assume that ${\rm Im}\,\alpha(t+i0)$ 
increases slowly enough at 
$t\to +\infty$ (for example, not faster than $Ct\ln^{-1-\epsilon}t$, 
$\epsilon>0$) so that the dispersion relations with 
not more than one subtraction take place, i.e. 
$$
\alpha(t)=\alpha_0+\frac{t}{\pi}\int_{t_{\rm T}}^{+\infty}
\frac{{\rm Im}\,\alpha(t'+i0)}{t'(t'-t)}dt'\,,
$$
and ${\rm Im}\,\alpha(t+i0)\ge 0$ at $t\ge t_{\rm T}>0$ 
we obtain
\begin{equation}
\label{gerg}
\frac{d^n\alpha(t)}{dt^n}>0\;\;(t<t_{\rm T},\;n=1,2,3,...).
\end{equation}
It is usually assumed that for the true Regge trajectories relations 
(\ref{gerg}) are fulfilled \cite{squires}.\footnote{Note that the 
amplitude depends explicitly on the functional form of the Regge 
trajectories only 
(not on their derivatives). So, for the phenomenological approximation 
it is sufficient to use some functions monotonous with one or several 
first derivatives.}

In \cite{godizov}, it was demonstrated that taking into 
account physical restrictions (\ref{meson}), (\ref{gluon}), 
(\ref{gerg}) allows us to describe the diffractive pattern of 
high-energy nucleon-nucleon scattering in the framework of 
the minimal phenomenological Regge-eikonal model with only 
three (nonlinear) Regge trajectories (soft Pomeron and two secondary 
Reggeons).\footnote{One could wonder if the nonlinearity of Regge 
trajectories related 
to high-$t$ behavior is so much important for diffractive scattering 
which occurs mainly at quite small transferred momenta. Isn't it 
more economical to deal with much more simple and convenient in many 
respects linear approximation $\alpha(t)\approx\alpha(0)+\alpha'(0)t$? 
In Ref. \cite{godizov}, we demonstrated that such an approximation 
is valid only in quite a narrow interval of $t$ (e.g. 
$|t|<0.3\,GeV^2$ for high-energy $pp$ scattering).} 

\begin{figure}
\epsfxsize=16.7cm\epsfysize=16.7cm\epsffile{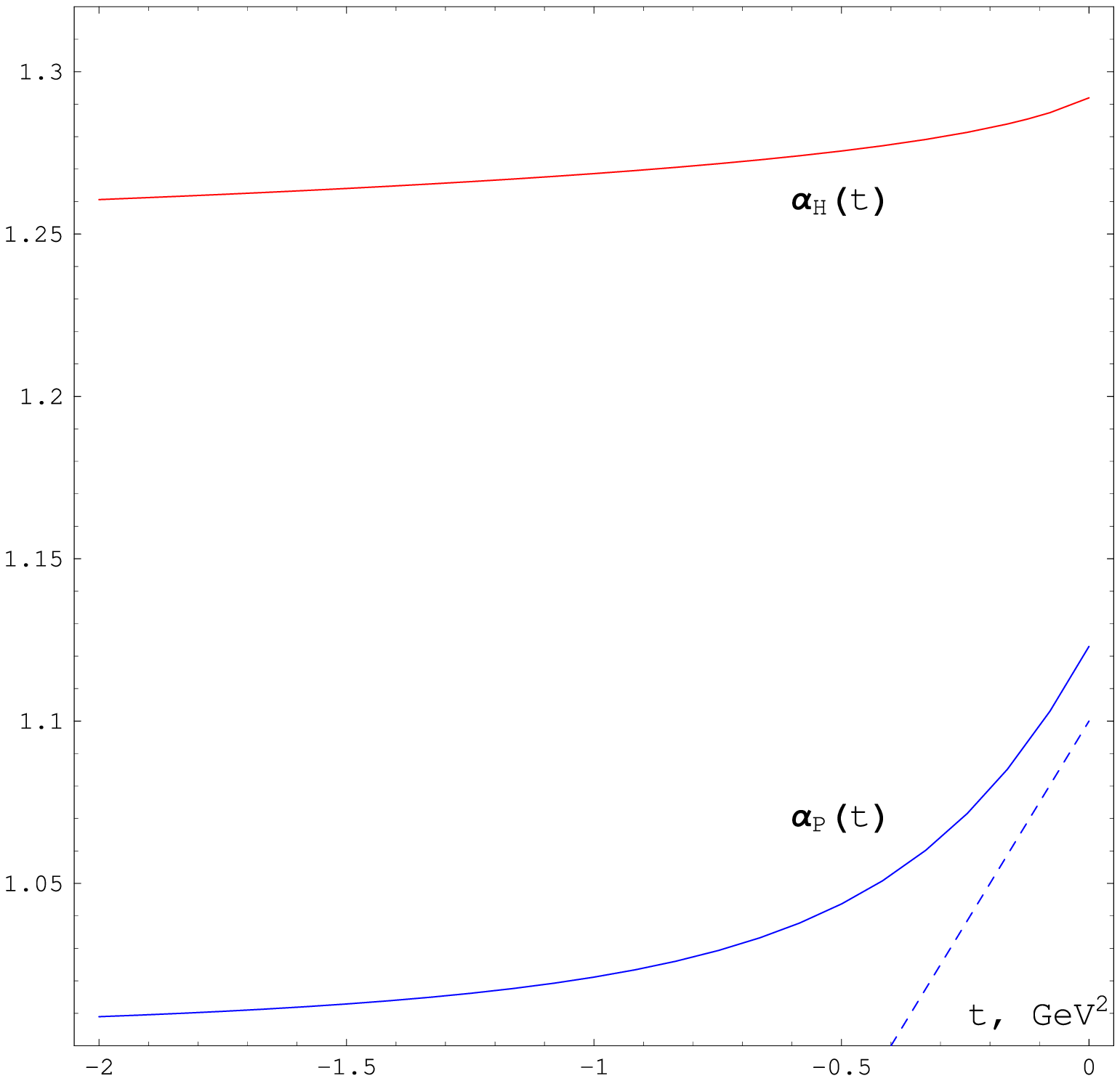}
\caption{Phenomenological approximations to soft ($\alpha_P(t)$) 
and hard ($\alpha_H(t)$) 
Pomeron trajectories obtained correspondingly by fitting the data 
on high-energy nucleon-nucleon scattering and photoproduction 
of $J/\psi$ (dashed line, 
$\alpha^{lin}_P(t)=1.1+0.25\,t$, is the linear soft Pomeron 
trajectory usually used in literature).}
\label{traV}
\end{figure}

Below we will show that such an approach also allows us to describe 
available data on high-energy ($W>30\,GeV$) exclusive vector meson 
photo- and electroproduction in the framework of the similar model, 
using only two $C$-even vacuum Regge trajectories, namely, 
soft and hard Pomerons (here we follow the pattern of 
Donnachie and Landshoff \cite{landshoff}). The soft Pomeron 
trajectory is the same as for the diffractive nucleon-nucleon scattering 
(in accordance with the requirement of universality of Regge 
trajectories) and the hard Pomeron trajectory, lying 
much higher than the soft one in the Euclidean 
domain (see Fig. \ref{traV}), has the asymptotic 
behaviour (\ref{hard}) (in accordance with the fact that at high 
transfers the high-energy scattering amplitude is dominated by 
two-gluon exchange). The ``invisibility'' of the hard Pomeron in the 
elastic diffraction may be explained by the fact that 
at available energies its contribution to the on shell eikonal is 
suppressed by the soft Pomeron due to the disparity 
in on shell residues $\beta_H(t)\ll\beta_P(t)$ (subscripts 
``$H$'' and ``$P$'' correspond to hard and soft) in the 
nonperturbative regime.

\section*{The model}

We will treat the processes of exclusive vector meson production 
from the standpoint of the vector-dominance model (VDM) 
\cite{sakurai} in which the incoming 
photon fluctuates into a virtual vector meson which, in turn, 
scatters from the target proton. According to the so-called 
hypothesis of $s$-channel helicity conservation (the 
adequacy of this approximation is confirmed by numerous experimental 
data on helicity effects \cite{jpsiele,phiele,rhoele}) 
the cross section of reaction 
$\gamma^*+p\to V+p$ is dominated by the nonflip helicity amplitude which 
can be represented in the form 
\begin{equation}
\label{genvdm}
T^\lambda_{\gamma^* p\to V p}(W^2,t,Q^2)=\sum_{V'}C_{V'}^\lambda(Q^2)
T^\lambda_{V'^* p\to V p}(W^2,t,Q^2)
\end{equation}
where the sum is taken over all neutral vector mesons. 
$C_{V'}^\lambda(Q^2)$ is the coefficient of vector dominance 
dependent on the type of vector meson $V'$, helicity $\lambda$ 
and virtuality $Q^2$ of the incoming photon. 
$T^\lambda_{V'^* p\to V p}(W^2,t,Q^2)$ is the 
diffractive hadronic amplitude with Regge-eikonal structure 
(the applicability of the Regge-eikonal approach to the hadronic 
reactions with off shell particles is grounded in \cite{petrov}). 

It is known that for small enough values of $t$ the 
diagonal coupling of vacuum Reggeons to 
hadrons is much stronger than the off diagonal coupling
(for example, elastic proton-proton scattering has a considerably 
larger cross section than diffractive excitation of N(1470) in the 
proton-proton collisions). So, in the high-energy range ($W>30\,GeV$) 
where vacuum exchanges exceed the nonvacuum ones 
(the main indication of such an exceeding is the manifest growth 
of integrated cross sections with the collision energy increase) 
the so-called ``diagonal approximation'' of (\ref{genvdm}) 
may be used 
\begin{equation}
\label{diavdm}
T^\lambda_{\gamma^* p\to V p}(W^2,t,Q^2)=
C_{V}^\lambda(Q^2)T_{V^* p\to V p}(W^2,t,Q^2)
\end{equation}
(conversely, for a consistent description of the data at $W<30\,GeV$ the 
contribution from nonvacuum Reggeons must be taken into account and 
approximation (\ref{diavdm}) becomes invalid). 
Here we must note that recent measurements by ZEUS Collaboration 
\cite{jpsiele,phiele,rhoele} 
did not reveal patent dependence on $t$ and $W$ of the ratio of 
differential cross sections for the cases of 
longitudinally and transversely polarized incoming photon and this 
is the reason why we have omitted the superscript $\lambda$ for 
$T_{V^* p\to V p}(W^2,t,Q^2)$ in (\ref{diavdm}). 
In other words, we presume that spin phenomena related to the 
dependence of $T_{V^* p\to V p}(W^2,t,Q^2)$ on helicities 
of incoming and outgoing particles are much more fine effects than
diffractive scattering itself and further deal only with 
quantities averaged over spin states. 

According to (\ref{diavdm}) and Regge theory \cite{squires} 
the pole contribution of any $C$-even Reggeon $R$ to the amplitude 
$T^\lambda_{\gamma^* p\to V p}(W^2,t,Q^2)$ has the form 
(the use of the designation $\delta$ for the pole contribution 
is caused by the subsequent use of the Regge-eikonal scheme; see below)
$$
C_{V}^\lambda(Q^2)\delta_R^{V^*}(W^2,t,Q^2) = 
C_{V}^\lambda(Q^2)\Gamma_R^{V^*V}(t,Q^2)\Gamma_R^{pp}(t)
\left(i+{\rm tg}\frac{\pi(\alpha_R(t)-1)}{2}\right)
\left(\frac{W^2}{W^2_0}\right)^{\alpha_R(t)}
$$
where $W_0\equiv 1\,GeV$, $\alpha_R(t)$ is the Reggeon trajectory, 
and $\Gamma_R^{V^*V}(t,Q^2)$, $\Gamma_R^{pp}(t)$ are the Reggeon 
form factors of participating particles. 

The VDM in its simplest form gives \cite{sakurai}
$$
C_{V}^{\pm 1}(Q^2) = \sqrt{\frac{3\Gamma_{V\to e^+e^-}}{\alpha_e M_{V}}}
\frac{M^2_{V}}{M^2_{V}+Q^2}\;\;,\;\;\;\;
\left(\frac{C_{V}^0(Q^2)}{C_{V}^{\pm 1}(Q^2)}\right)^2=\frac{Q^2}{M_V^2}
$$
where $M_{V}$ is the vector meson mass, $\alpha_e=\frac{1}{137}$ 
is the electromagnetic coupling, and $\Gamma_{V\to e^+e^-}$ is the 
width of the vector meson decay to the electron-positron pair.
However, these relations contradict both the experimental data 
on $Q^2$ dependence of the ratio $R(Q^2)=\frac{\sigma^0_{\gamma^* p\to V p}}
{\sigma^{\pm 1}_{\gamma^* p\to V p}}\approx
\left(\frac{C_{V}^0(Q^2)}{C_{V}^{\pm 1}(Q^2)}\right)^2$ 
\cite{jpsiele,phiele,rhoele} 
and the dimensional counting rules \cite{matveev} according to which functions 
$Q^2\Gamma_R^{V^*V}(t,Q^2)$, $Q^2C_{V}^\lambda(Q^2)\Gamma_R^{V^*V}(t,Q^2)$, 
and, hence, $C_{V}^\lambda(Q^2)$ must exhibit moderate 
(nonpowerlike) $Q^2$ behavior. 

To resolve this conflict we have to modify the VDM coefficients 
\begin{equation}
\label{vdmcoe}
C_{V}^\lambda(Q^2) = \sqrt{\frac{3\Gamma_{V\to e^+e^-}}{\alpha_e M_{V}}}
\frac{M^2_{V}}{M^2_{V}+Q^2}F^\lambda_V(Q^2)
\end{equation}
and thus 
$$
C_{V}^\lambda(Q^2)\delta_R^{V^*}(W^2,t,Q^2) = 
\sqrt{\frac{3\Gamma_{V\to e^+e^-}}{\alpha_e M_{V}}}
\frac{M^2_{V}}{M^2_{V}+Q^2}\times
$$
$$
\times F^\lambda_V(Q^2)
\Gamma_R^{V^*V}(t,Q^2)\Gamma_R^{pp}(t)
\left(i+{\rm tg}\frac{\pi(\alpha_R(t)-1)}{2}\right)
\left(\frac{W^2}{W^2_0}\right)^{\alpha_R(t)}
$$
where, for any virtual photon spin 
state $\lambda$, $F^\lambda_V(-M_V^2)=1$, and function 
$F^\lambda_V(Q^2)\Gamma_R^{V^*V}(t,Q^2)$ exhibits nonsteep 
$Q^2$ behavior. 

Further on, we will concentrate on the diffractive 
pattern ($t$ dependence) of exclusive vector meson production and 
its evolution with collision energy ($W$ dependence). 
Since the second factor in the right-hand side of (\ref{diavdm}) 
nearly does not depend on $\lambda$ we may 
consider only cross sections averaged over helicities of 
external particles. The averaged pole contribution of a Reggeon 
$R$ to the amplitude takes the form 
\begin{equation}
\label{poleconz}
\bar C_{V}(Q^2)\delta_R^{V^*}(W^2,t,Q^2) = 
\sqrt{\frac{3\Gamma_{V\to e^+e^-}}{\alpha_e M_{V}}}
\frac{M^2_{V}}{M^2_{V}+Q^2}\times
\end{equation}
$$
\times\bar\Gamma_R^{V^*V}(t,Q^2)\Gamma_R^{pp}(t)
\left(i+{\rm tg}\frac{\pi(\alpha_R(t)-1)}{2}\right)
\left(\frac{W^2}{W^2_0}\right)^{\alpha_R(t)}
$$
where 
$$
\bar C_{V}(Q^2)\equiv
\sqrt{\frac{3\Gamma_{V\to e^+e^-}}{\alpha_e M_{V}}}
\frac{M^2_{V}}{M^2_{V}+Q^2}\bar F_{V}(Q^2)\,,\;\;\;
\bar F_{V}(Q^2)\equiv\sqrt{\frac{1}{3}
\sum_\lambda(F_{V}^\lambda(Q^2))^2}\,,
$$
$$
\bar\Gamma_R^{V^*V}(t,Q^2)\equiv \bar F_{V}(Q^2)\Gamma_R^{V^*V}(t,Q^2)\,,
\;\;\;\bar\Gamma_R^{V^*V}(t,-M_V^2)\equiv \Gamma_R^{VV}(t)\,.
$$
In a well-known paper \cite{gellmann}, it was argued that 
$\bar F_{V}(Q^2)=1+\frac{Q^2}{m_{V0}^2}$ where $m_{V0}$ is the 
vector meson ``bare'' mass. At infinite values of $m_{V0}$ the 
usual form of the current-field relation takes place. An interesting 
feature of this variant is a possibility to extend the VDM 
applicability for arbitrarily high $Q^2$. We do not fix the 
concrete functional form of $\bar F_{V}(Q^2)$ (it does not help to 
determine $Q^2$ behavior of $\bar\Gamma_R^{V^*V}(t,Q^2)$ because 
of the absence of detailed 
information on $Q^2$ dependence of $\Gamma_R^{V^*V}(t,Q^2)$). 
According to the dimensional counting rules \cite{matveev} 
$\Gamma_R^{V^*V}(t,Q^2)\sim Q^{-2}$ at high $Q^2$ 
(we point out that such a behavior implies the violation of Bjorken
scaling \cite{londergan}), 
so $\bar\Gamma_R^{V^*V}(t,Q^2)$ must exhibit slow evolution 
with $Q^2$. 

After these general remarks we can proceed to construct the 
scattering amplitude. 
It is well known that for soft reactions such as elastic 
nucleon-nucleon scattering or 
light vector meson photoproduction the introduction of only one trajectory 
with intercept higher than unity is necessary (the soft Pomeron). 
But in the processes 
with a hard scale (for example, photon virtuality or/and heavy vector meson 
mass) the rise of the cross sections with the collision energy becomes 
noticeably faster than in the soft ones and so there emerges the question 
if there exists one or several extra Reggeons lying higher than 
the soft Pomeron but with entirely suppressed residue(s) in the 
nonperturbative regime \cite{prokudin}.\footnote{Introducing extra trajectories 
may seem a violation of the Occam razor principle. Actually, 
we did try a version with 
a single Pomeron but in this case one needs negative and growing with 
$Q^2$ residues for secondary trajectories ($f_2$ Reggeon etc.). The 
last circumstance looks too counter-intuitive and a hard Pomeron seems 
inevitable.} 

Below we will use the minimal 
phenomenological scheme based on the idea of Donnachie and 
Landshoff \cite{landshoff,donnachie} that only one extra Reggeon with 
the intercept much higher than the soft Pomeron's one is needed. 
The difference between our realization of this idea and the 
Donnachie-Landshoff model lies in a choice of the functional form of 
Regge trajectories since the authors \cite{landshoff,donnachie} insist on the 
strict linearity of the trajectories while we use essentially nonlinear 
parametrizations with asymptotic behavior following from QCD (cf. the 
footnote 2).

We assume that at collision energies 
$W>30\,GeV$ we may neglect contributions from secondary Reggeons. 
In this case the minimal vacuum exchange off shell eikonal (the sum of 
one-Reggeon-exchange amplitudes \cite{arnold}) takes the form 
$$
\bar C_{V}(Q^2)\delta^{V^*}(W^2,t,Q^2)\equiv 
\delta^*(W^2,t,Q^2)
= \delta^*_P(W^2,t,Q^2)+
\delta^*_H(W^2,t,Q^2)=
$$
\begin{equation}
\label{eikphenz}
=\sqrt{\frac{3\Gamma_{V\to e^+e^-}}{\alpha_e M_{V}}}
\frac{M^2_{V}}{M^2_{V}+Q^2}
\left[\left(i+{\rm tg}\frac{\pi(\alpha_P(t)-1)}{2}\right)
\bar\Gamma_P^{V^*V}(t,Q^2)\Gamma_P^{pp}(t)
\left(\frac{W^2}{W^2_0}\right)^{\alpha_P(t)}+\right.
\end{equation}
$$
+\left.\left(i+{\rm tg}\frac{\pi(\alpha_H(t)-1)}{2}\right)
\bar\beta^{V^*}_H(t,Q^2)
\left(\frac{W^2}{W^2_0}\right)^{\alpha_H(t)}\right]\,,
$$
where $\alpha_P(t)$, $\alpha_H(t)$ are Regge trajectories of 
the soft and hard Pomerons, $\bar\Gamma_P^{V^*V}(t,Q^2)$ 
and $\Gamma_P^{pp}(t)$ are the soft Pomeron form factors of 
outgoing vector meson and proton, and 
$\bar\beta^{V^*}_H(t,Q^2)\equiv
\bar\Gamma_H^{V^*V}(t,Q^2)\Gamma_H^{pp}(t)$ is the 
Regge residue of the hard Pomeron. 
Factors $\bar\Gamma_H^{V^*V}(t,Q^2)$ and $\Gamma_H^{pp}(t)$ are 
absorbed into an overall hard residue $\bar\beta^{V^*}_H(t,Q^2)$ 
since the hard Pomeron gives no noticeable contribution to 
the eikonal of nucleon-nucleon scattering at accessible energies and so 
we can not separately 
fix the form factor $\Gamma_H^{pp}(t)$ from the data on 
high-energy proton-(anti)proton diffraction. 

The corresponding on-shell eikonal of elastic $Vp$ scattering is of the form: 
\begin{equation}
\label{eikphen}
\delta^V(W^2,t) = 
\left(i+{\rm tg}\frac{\pi(\alpha_P(t)-1)}{2}\right)
\Gamma_P^{VV}(t)\Gamma_P^{pp}(t)
\left(\frac{W^2}{W^2_0}\right)^{\alpha_P(t)}+
\end{equation}
$$
+\left(i+{\rm tg}\frac{\pi(\alpha_H(t)-1)}{2}\right)
\beta^V_H(t)\left(\frac{W^2}{W^2_0}\right)^{\alpha_H(t)}\,.
$$

Since Regge trajectories are universal functions of one argument, 
i.e. depend neither on the type of the process in which the 
corresponding Reggeons contribute nor on the virtualities of the 
incoming particles, we will use, for the description 
of exclusive diffractive production of vector mesons, 
the same phenomenological expressions for the soft 
Pomeron trajectory and the corresponding form factor of the proton 
as we did for high-energy nucleon-nucleon scattering \cite{godizov}: 
\begin{equation}
\label{pomeron}
\alpha_P(t) = 1+p_1\left[1-p_2\,t\left({\rm arctg}(p_3-p_2\,t)
                             -\frac{\pi}{2}\right)\right]\,,
\end{equation}
$$
\beta_P^{pp}(t)=(\Gamma_P^{pp}(t))^2 = B_Pe^{b_P\,t}
(1+d_1\,t+d_2\,t^2+d_3\,t^3+d_4\,t^4)
$$
(the values of free parameters obtained by fitting 
the data on proton-(anti)proton angular distributions are 
represented in Table \ref{tab1}).

\begin{table}[h]
\begin{center}
\begin{tabular}{|l|l|l|l|}
\hline
$p_1$ & $0.123$           & $d_1$ & $0.43\,GeV^{-2}$  \\
\hline
$p_2$ & $1.58\,GeV^{-2}$  & $d_2$ & $0.39\,GeV^{-4}$  \\
\hline
$p_3$ & $0.15$            & $d_3$ & $0.051\,GeV^{-6}$ \\
\hline
$B_P$ & $43.5$            & $d_4$ & $0.035\,GeV^{-8}$ \\
\hline
$b_P$ & $2.4\,GeV^{-2}$   & &\\
\hline
\end{tabular}
\end{center}
\caption{Parameters for $\alpha_P(t)$ 
and $\Gamma_P^{pp}(t)$ obtained by fitting the data 
on proton-(anti)proton angular distributions.}
\label{tab1}
\end{table}

The hard Pomeron trajectory 
is chosen in the form similar to the one used in \cite{brodsky} 
for the $\rho$ Reggeon 
\begin{equation}
\label{hardpom}
\alpha_H(t) = 1+\frac{1}{A_H+
[\frac{12\,\ln 2}{\pi}\alpha_s(\sqrt{-t+c_H})]^{-1}}
\end{equation}
where 
$$
\alpha_s(\mu) \equiv \frac{4\pi}{11-\frac{2}{3}n_f}
\left(\frac{1}{\ln\frac{\mu^2}{\Lambda^2}}
+\frac{1}{1-\frac{\mu^2}{\Lambda^2}}\right)
$$
is the so-called (one-loop) analytical 
QCD effective coupling constant \cite{solovtsov}, 
$n_f = 3$ is the number of quark flavors taken into account, 
$\Lambda = \Lambda^{(3)} = 0.346\,GeV$ is the QCD dimensional parameter 
(the value was taken from \cite{bethke}) 
and $A_H$ and $c_H$ are free parameters. 
Note that our approximations to the soft and hard Pomerons 
satisfy asymptotic relations (\ref{gluon}) and (\ref{hard}) 
thus corresponding to multi- and two-gluon exchanges at high transferred 
momenta. 

Form-factor $\bar\Gamma_P^{V^*V}(t,Q^2)$ and residue 
$\bar\beta^{V^*}_H(t,Q^2)$ are assumed to have approximately 
an exponential form for any value of $Q^2$ and small enough 
values of $t$ 
\begin{equation}
\label{resid}
\bar\Gamma_P^{V^*V}(t,Q^2) = B^V_P(Q^2)e^{b^V_P(Q^2)\,t},\;\;\;\;
\bar\beta^{V^*}_H(t,Q^2) = B^V_H(Q^2)e^{b^V_H(Q^2)\,t}\;\;\;\;
(V=\rho,\phi,J/\psi).
\end{equation}
Here $B^V_P(Q^2)$ characterizes the effective intensity 
of the interaction between vector meson $V$ and soft Pomeron 
and $b^V_P(Q^2)$ is associated with the corresponding effective radius 
of interaction. 

To obtain angular distributions we substitute (\ref{pomeron}), 
(\ref{hardpom}), (\ref{resid}) into (\ref{eikphenz}), (\ref{eikphen})
and proceed via Fourier-Bessel transformation 
\begin{equation}
\label{fourier1}
\delta^*(W^2,b,Q^2) = 
\frac{1}{16\pi W^2}\int_0^{\infty}d(-t)J_0(b\sqrt{-t})
\delta^*(W^2,t,Q^2)\,,
\end{equation}
$$
\delta^V(W^2,b) = 
\frac{1}{16\pi W^2}\int_0^{\infty}d(-t)J_0(b\sqrt{-t})\delta^V(W^2,t)
$$
to the impact parameter representation. 

For obtaining the full amplitude 
we take use of an extended (off shell) eikonal representation \cite{petrov} 
\begin{equation}
\label{eikrepr}
T_{\gamma^* p\to V p}(W^2,b,Q^2) = 
\frac{\delta^*(W^2,b,Q^2)}
{\delta^V(W^2,b)}T_{V p\to V p}(W^2,b) = 
\end{equation}
$$
=\frac{\delta^*(W^2,b,Q^2)}
{\delta^V(W^2,b)}\frac{e^{2i\delta^V(W^2,b)}-1}{2i} = 
\delta^*(W^2,b,Q^2)+i\delta^*(W^2,b,Q^2)\delta^V(W^2,b)+...
$$
(here $T_{V p\to V p}(W^2,b)=\frac{e^{2i\delta^V(W^2,b)}-1}{2i}$ is 
the ``eikonalized'' (unitarized) amplitude of elastic $Vp$-scattering). 
The inverse Fourier-Bessel transformation 
\begin{equation}
\label{fourier2}
T_{\gamma^* p\to V p}(W^2,t,Q^2) = 4\pi W^2\int_0^{\infty}db^2J_0(b\sqrt{-t})
T_{\gamma^* p\to V p}(W^2,b,Q^2)
\end{equation}
gives the full amplitude 
in the momentum representation (during numerical calculating integrals from 
(\ref{fourier1}), (\ref{fourier2}) we approximate upper limits of integration 
by $8\,GeV^2$ and 
$250\,GeV^{-2}\approx(3\,fm)^2$ correspondingly) and then is used 
for the differential cross section 
\begin{equation}
\label{diffsech}
\frac{d\sigma_{\gamma^* p\to V p}}{dt} = 
\frac{|T_{\gamma^* p\to V p}(W^2,t,Q^2)|^2}{16\pi W^4}\,.
\end{equation}

A different way of unitarizing off shell vector meson production 
amplitude was exploited in \cite{troshin}.

\section*{The description of experimental data}

Turn to the description of experimental data on exclusive 
diffractive production of vector mesons 
(the description of available angular distributions 
in the framework of another 
phenomenological approaches without ``unitarization'' can be found in 
\cite{donnachie,predazzi,fiore,motyka,soyez}). 
At the very start 
we must note that because of the absence of high-energy data 
on elastic $Vp$ scattering we have to make an assumption that 
$\Gamma_P^{VV}(t)=\bar\Gamma_P^{V^*V}(t,-M_V^2)
\approx\bar\Gamma_P^{V^*V}(t,0)$ and 
$\beta^{V}_H(t)=\bar\beta^{V^*}_H(t,-M_V^2)
\approx\bar\beta^{V^*}_H(t,0)$. Below, 
it will be shown that corresponding quantities change very 
slowly with $Q^2$ (in accordance with the dimensional counting rules 
\cite{matveev}) and, so, such an 
approximation seems to be justified from the phenomenological point 
of view. In other words, for all reactions 
we will determine, at first, the on-shell 
eikonal $\delta^V(W^2,t)$ from the data on photoproduction 
and then proceed to electroproduction of the considered 
vector meson. In accordance with experimental data the 
cross sections of exclusive production are obtained by 
integration over $0<-t<0.6\,GeV^2$ for light mesons, over 
$0<-t<1.25\,GeV^2$ for the photoproduction of $J/\psi$, and 
over $0<-t<1.0\,GeV^2$ for the electroproduction of $J/\psi$.

We start from the consideration of exclusive production of 
$\phi$ meson \cite{phipho,phiele}. Since the main set of 
corresponding data on angular 
distributions and integrated cross sections is concentrated 
in the region of not very high energies $W<140\,GeV$ and 
photon virtualities $Q^2<14\,GeV^2$, and experimental errors 
are rather large we can significantly simplify our 
phenomenological model by neglecting the contribution of 
the hard Pomeron (i.e. we put 
$\bar\beta^{\phi^*}_H(t,Q^2)\approx 0$) not only at 
$Q^2\approx 0$ but also at 
all other values. Moreover, for $Q^2>2\,GeV^2$ we 
will neglect the $t$ dependence of 
$\bar\Gamma_P^{\phi^*\phi}(t,Q^2)$ by putting 
$b^\phi_P(Q^2)\approx 0$ (such an approximation implies that at 
$Q^2>2\,GeV^2$ soft Pomeron interacts with pointlike 
objects inside $\phi$ (in accord with the parton model) 
and $t$ dependence of the corresponding Regge 
residue is entirely determined by the form factor 
$\Gamma_P^{pp}(t)$). So for all nonzero values of $Q^2$ 
we will have only one free parameter, $B^\phi_P(Q^2)$, 
since $\alpha_P(t)$ and 
$\Gamma_P^{pp}(t)$ are fixed from the data on nucleon-nucleon 
scattering. 

\begin{figure}
\epsfxsize=16.7cm\epsfysize=16.7cm\epsffile{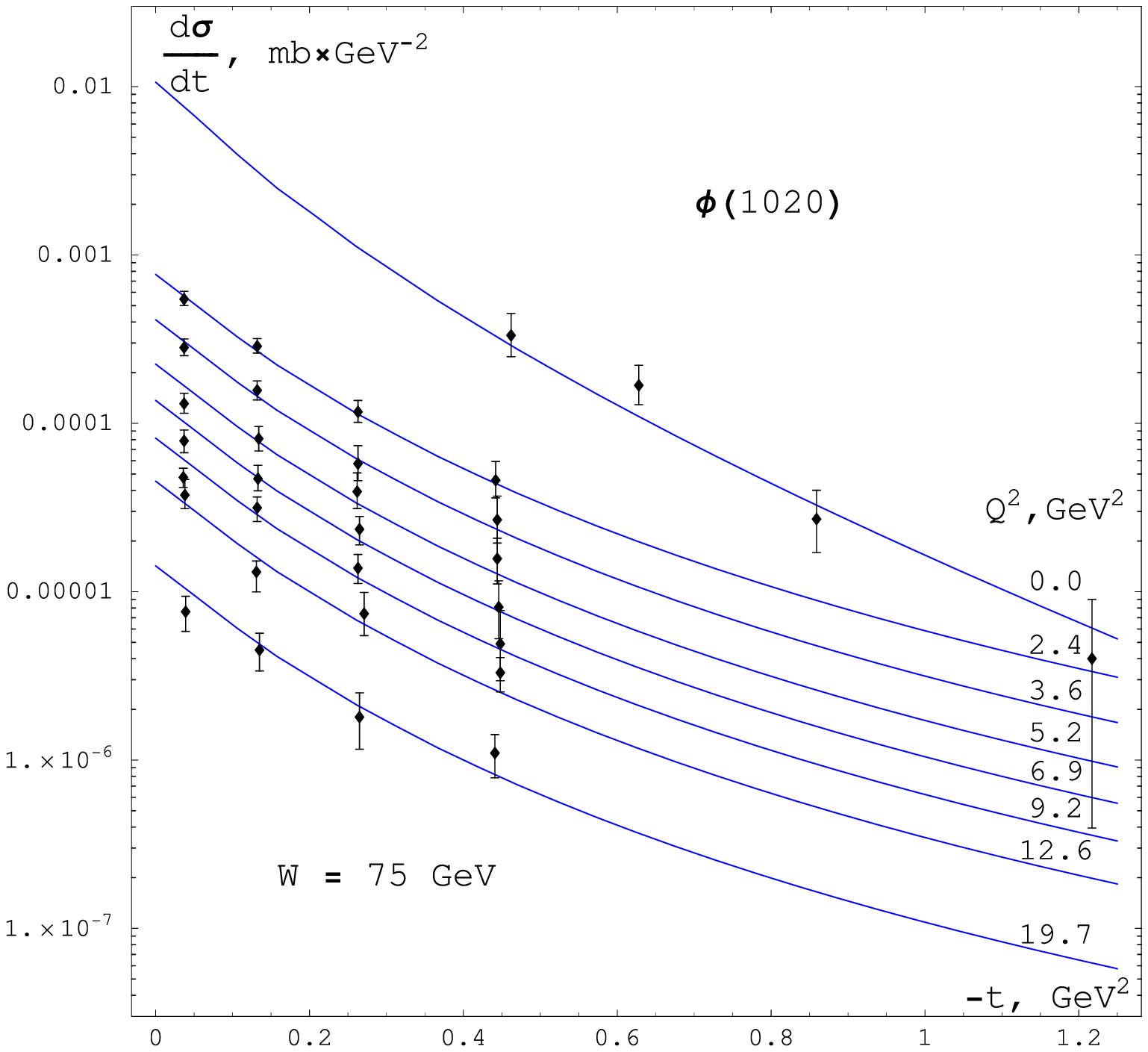}
\caption{Differential cross sections for exclusive 
$\phi$-meson electroproduction at collision energy 
$W=75\,GeV$ and different values of the incoming 
photon virtuality.}
\label{diffphi}
\end{figure}

\begin{figure}
\epsfxsize=16.7cm\epsfysize=16.7cm\epsffile{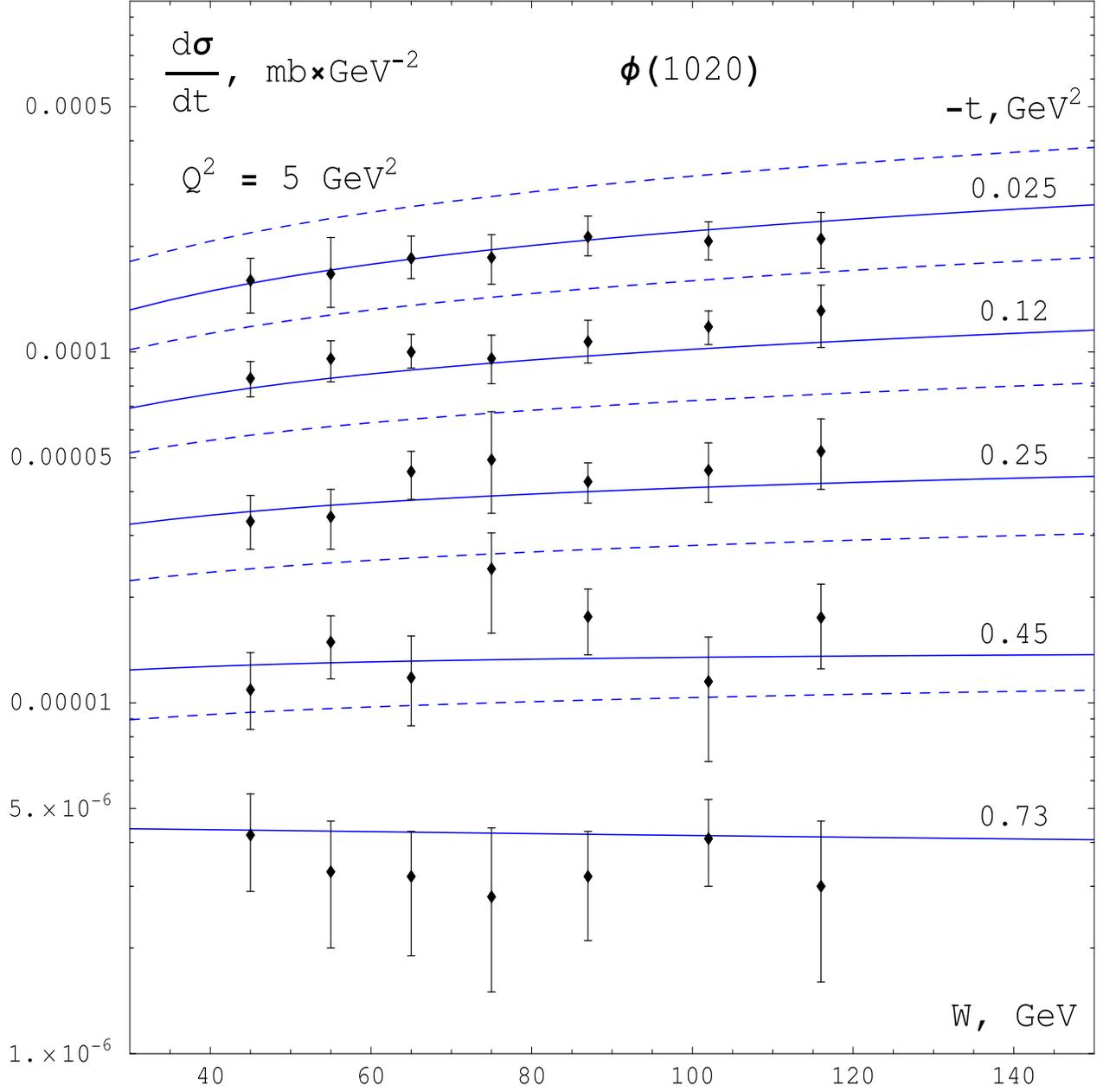}
\caption{Differential cross sections for exclusive 
$\phi$-meson electroproduction at $Q^2=5\,GeV^2$ and 
different values of transferred momentum squared as 
function of collision energy (1 free parameter, 35 points, 
$\chi^2=17.5$). Dashed lines correspond to 
cross sections in the Born approximation.}
\label{phi5}
\end{figure}

\begin{figure}
\epsfxsize=16.7cm\epsfysize=16.7cm\epsffile{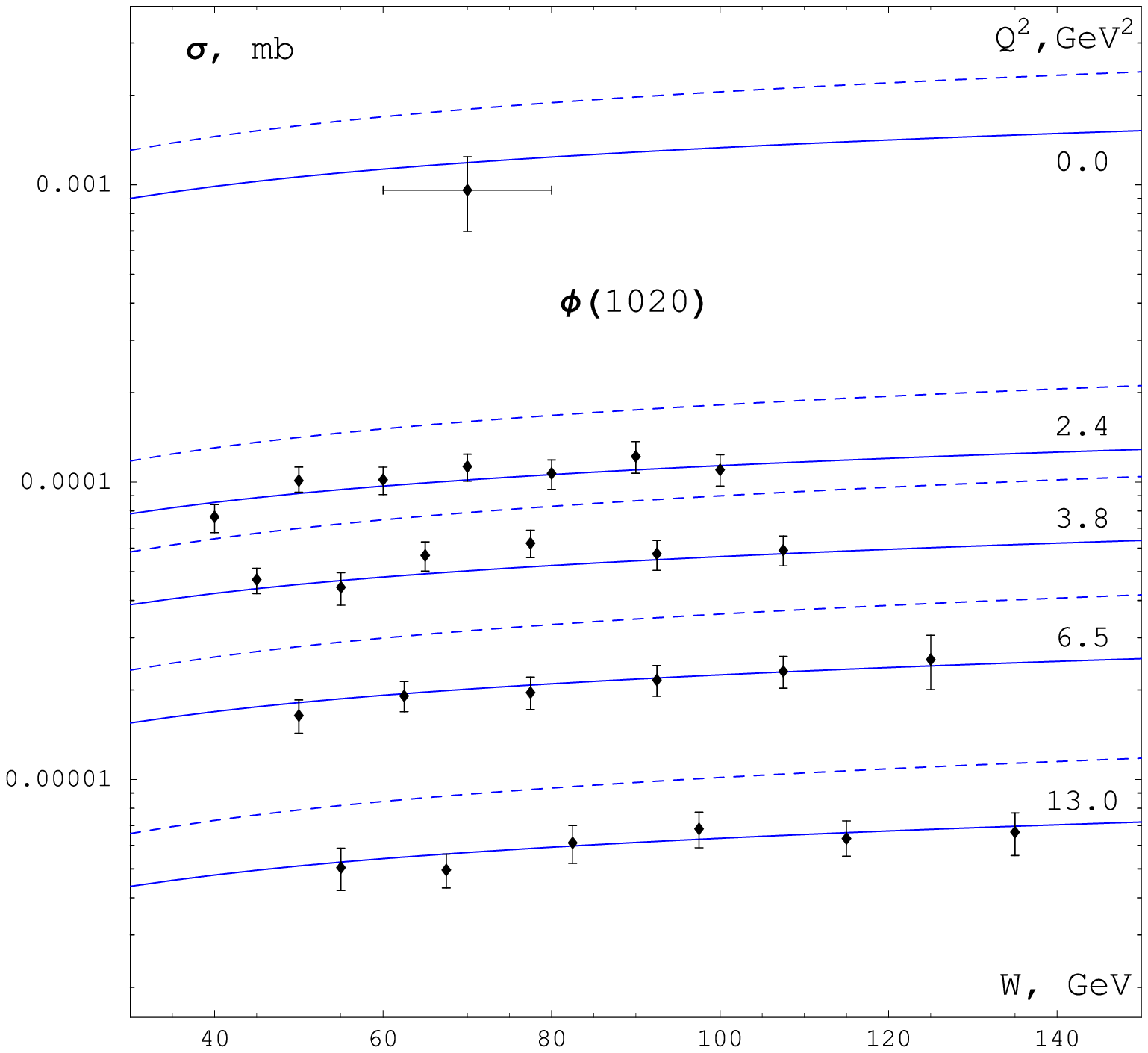}
\caption{Integrated cross sections for exclusive $\phi$-meson 
electroproduction at different values of the incoming 
photon virtuality as functions of collision energy. Dashed 
lines correspond to 
cross sections in the Born approximation.}
\label{elaphi}
\end{figure}

The results of fitting the data are represented in 
Table \ref{tab2} and Figs. \ref{diffphi}---\ref{elaphi}. 
These figures show that our quite rough 
approximation provides the data description of a 
satisfactory quality especially for low values of $Q^2$ 
(see Fig. \ref{phi5}), pointing to the fact that the soft 
Pomeron dominates even at nonzero photon virtualities 
and the contribution from the hard Pomeron lies 
within experimental errors. Figures \ref{phi5}, \ref{elaphi} 
also demonstrate that, in general, neglecting absorptive 
corrections (i.e. Regge cuts contribution) is inadmissible. 
\begin{table}[h]
\begin{center}
\begin{tabular}{|l|l|l|l|l|l|l|l|l|}
\hline
$Q^2$, $GeV^2$ & 0.0 & 2.4 & 3.6 & 5.0 & 6.5 & 9.2 & 13.0 & 19.7 \\
\hline
$B^\phi_P(Q^2)$ & 3.1 & 2.8 & 2.77 & 2.75 & 2.73 & 2.72 & 2.7 & 2.3  \\
\hline
$b^\phi_P(Q^2)$, $GeV^{-2}$ & 0.6 & \multicolumn{7}{|c|}{0 (fixed)}  \\
\hline
$<|t|>_{W=75\,GeV}$, $GeV^2$ & 0.13 & \multicolumn{7}{|c|}{0.20}  \\
\hline
\end{tabular}
\end{center}
\caption{Results of fitting 
the data on reaction $\gamma^* + p\to \phi + p$.}
\label{tab2}
\end{table}

In accordance with remarks on 
the $Q^2$-dependence of $\bar\Gamma_P^{V^*V}(t,Q^2)$ in the 
previous section, $B^\phi_P(Q^2)$ changes rather slowly 
with $Q^2$. Also in Table \ref{tab2} the 
average value of $|t|$ at $W = 75\,GeV$ is exhibited. 

\begin{figure}
\epsfxsize=16.7cm\epsfysize=16.7cm\epsffile{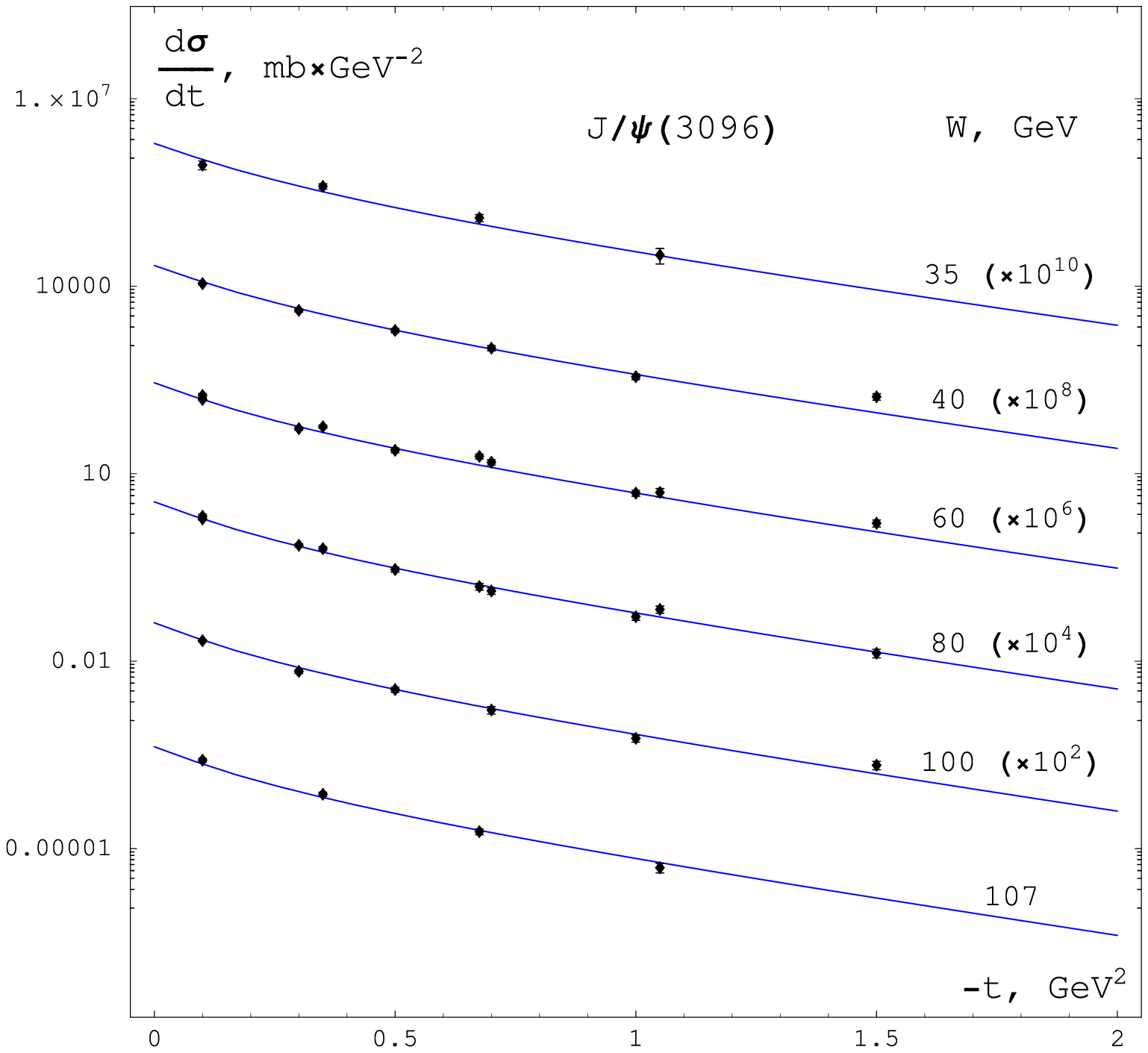}
\caption{Differential cross sections for exclusive 
$J/\psi$-meson 
photoproduction at different values of collision energy.}
\label{jps01}
\end{figure}

\begin{figure}
\epsfxsize=16.7cm\epsfysize=16.7cm\epsffile{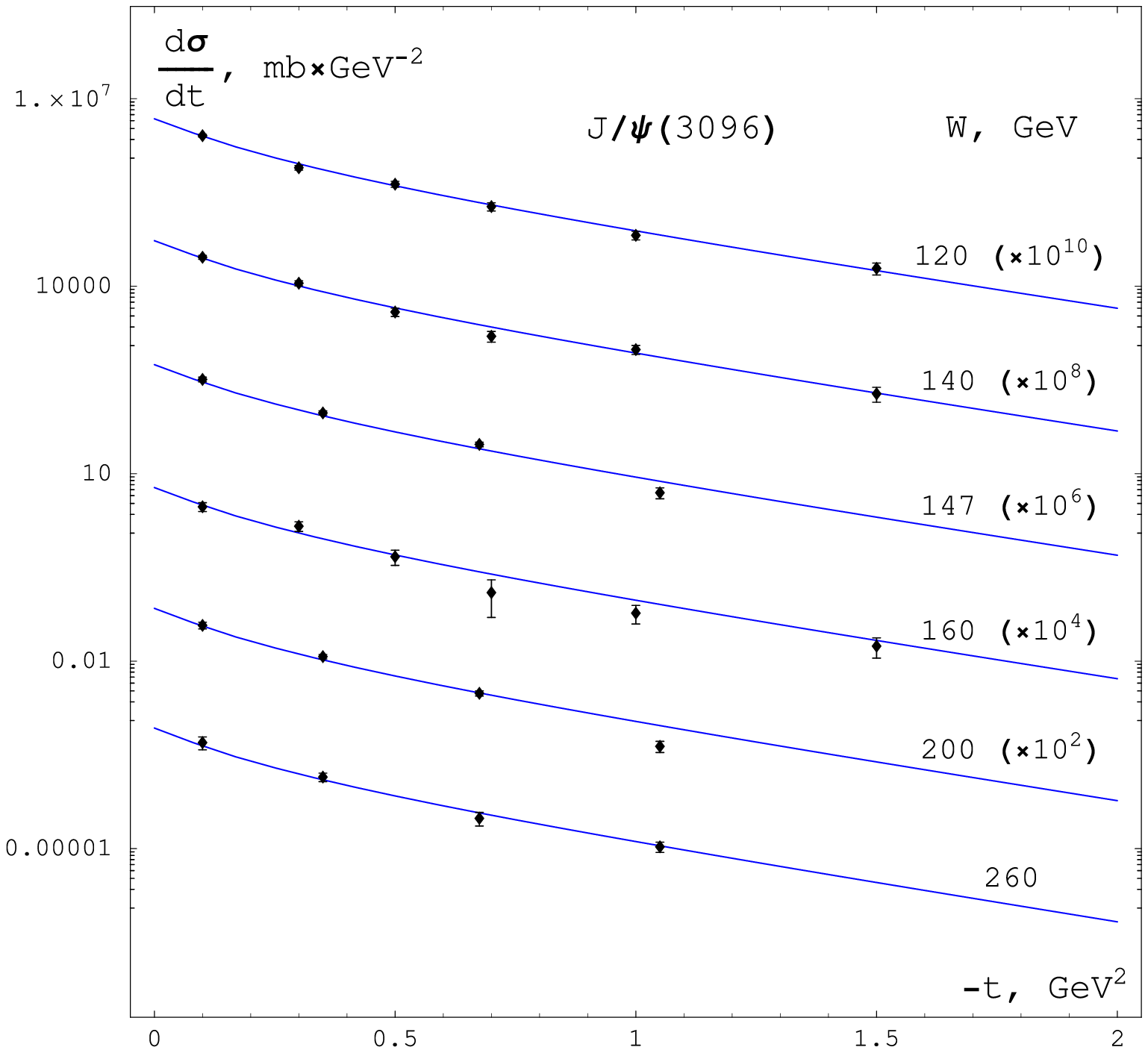}
\caption{Differential cross sections for exclusive 
$J/\psi$-meson 
photoproduction at different values of collision energy.}
\label{jps02}
\end{figure}

\begin{figure}
\epsfxsize=16.7cm\epsfysize=16.7cm\epsffile{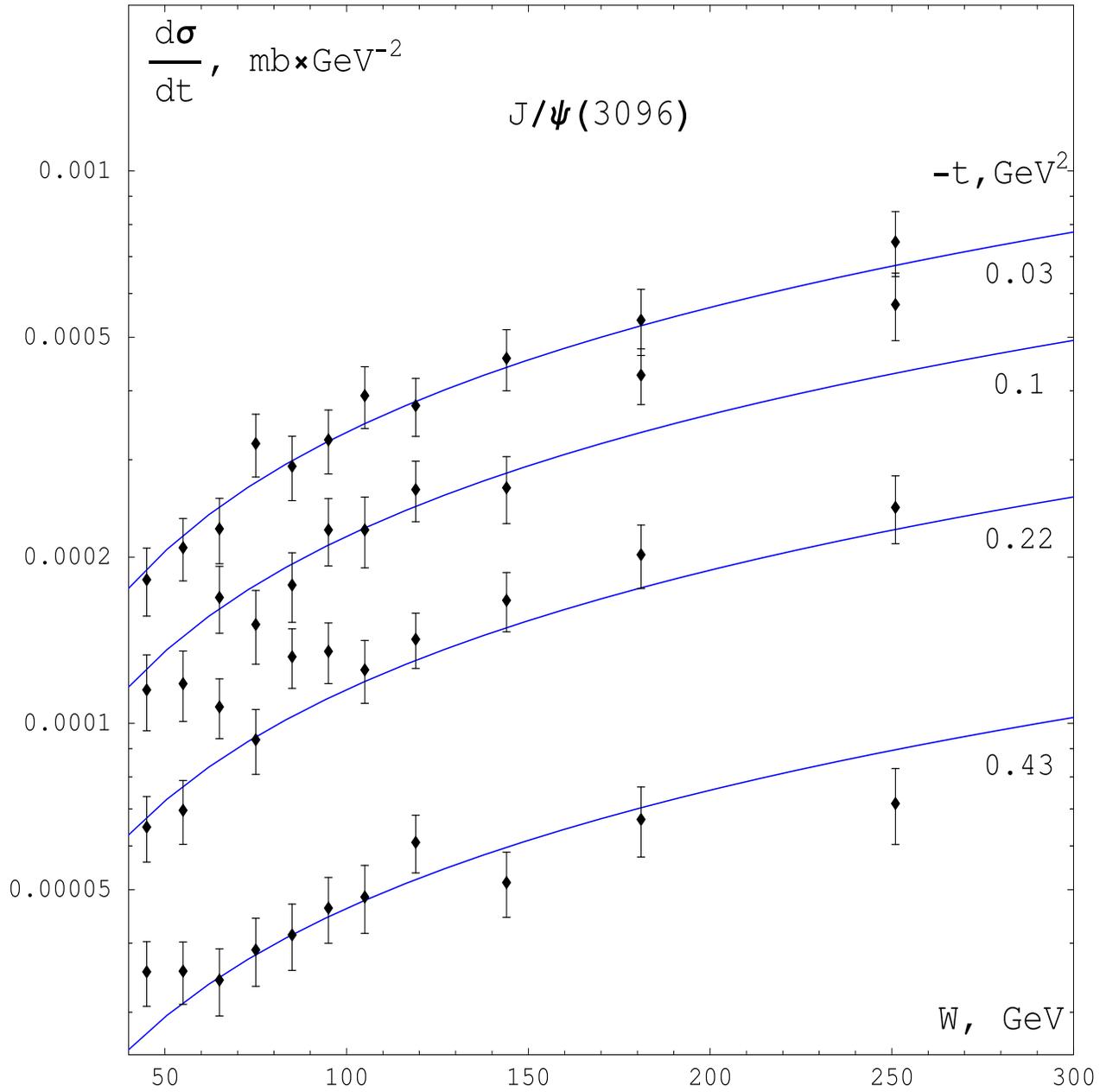}
\caption{Differential cross sections for exclusive 
$J/\psi$-meson photoproduction at different values 
of transferred momentum squared as function of collision 
energy.}
\label{jps03}
\end{figure}

\begin{figure}
\epsfxsize=16.7cm\epsfysize=16.7cm\epsffile{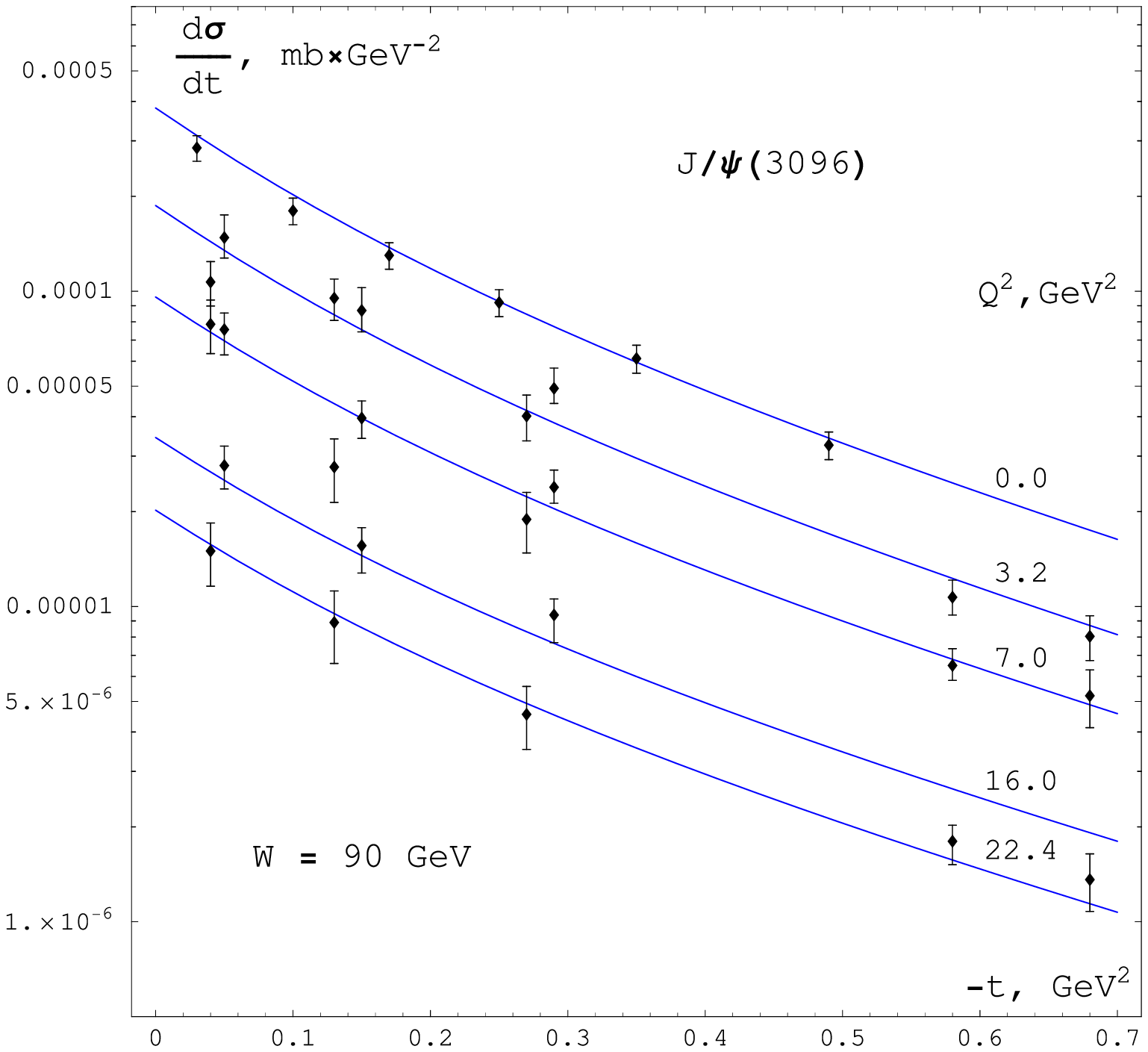}
\caption{Differential cross sections for exclusive 
$J/\psi$-meson electroproduction at collision energy 
$W=90\,GeV$ and different values of the incoming 
photon virtuality.}
\label{diffjps}
\end{figure}

\begin{figure}
\epsfxsize=16.7cm\epsfysize=16.7cm\epsffile{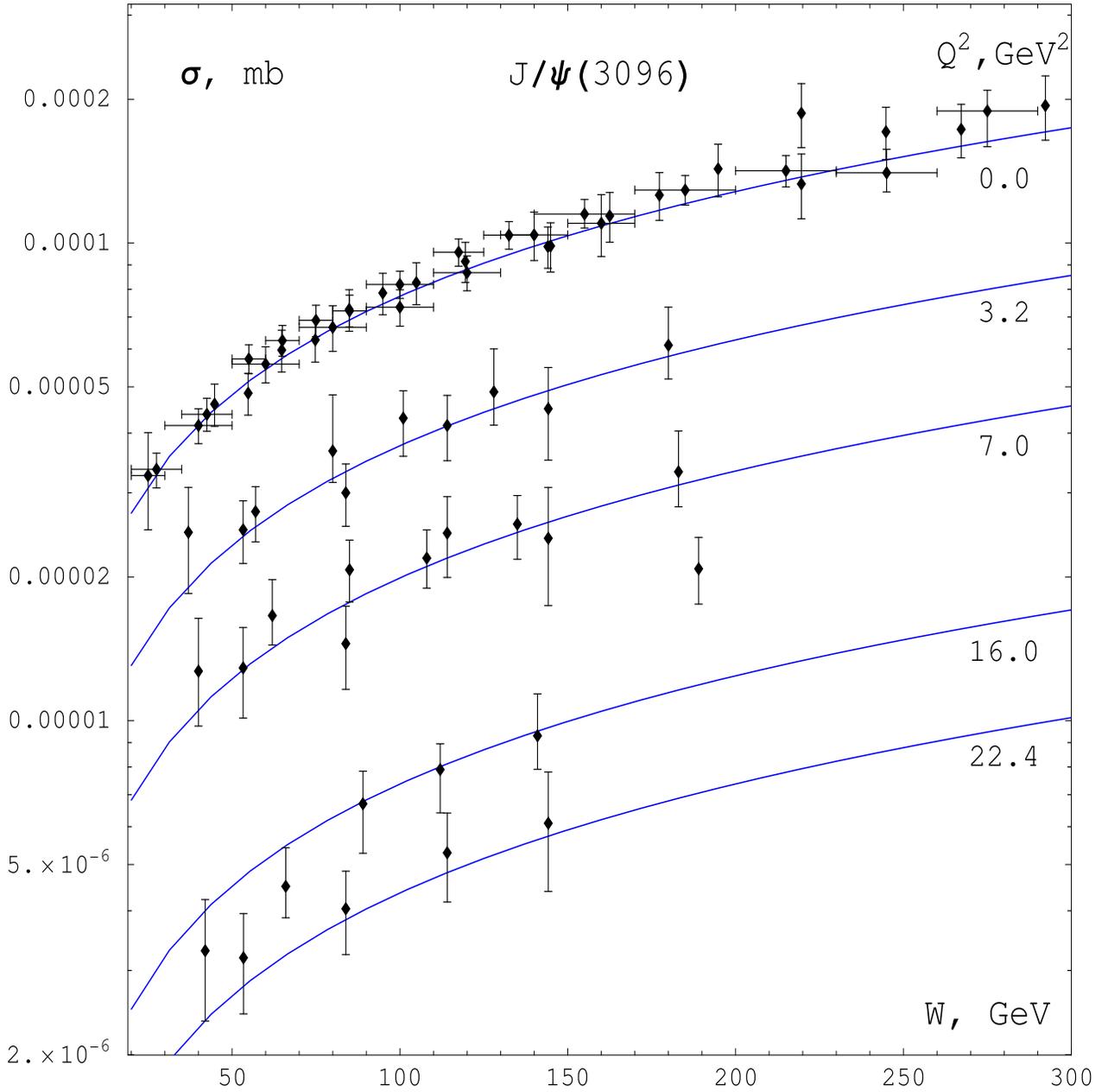}
\caption{Integrated cross sections for exclusive 
$J/\psi$-meson 
electroproduction at different values of the incoming 
photon virtuality as functions of collision energy.}
\label{elajps}
\end{figure}

The amount and the quality of data on $J/\psi$ electroproduction 
\cite{jpsipho,jpsiele} allow 
us to neglect neither the hard Pomeron's contribution nor the 
$t$ dependence of $\bar\Gamma_P^{{J/\psi}^*{J/\psi}}(t,Q^2)$. The results 
of fitting are represented in 
Tab. \ref{tab3} and Figs. \ref{jps01}---\ref{elajps}. For the 
$J/\psi$ photoproduction we obtain $\chi^2=194$ over 114 experimental 
points. The evolution of $<|t|>$ (which is related to the transverse 
interaction radius, $R^2\sim <|t|>^{-1}$) at fixed $W$ ($W = 90\,GeV$) 
is rather slow at all values of $Q^2$ 
due to the fact that even at $Q^2\approx 0$ the perturbative regime 
takes place because of presence of the hard scale $M_{J/\psi}$.

\begin{table}[h]
\begin{center}
\begin{tabular}{|l|l|l|l|l|l|}
\hline
$Q^2$, $GeV^2$ & 0.0 & 3.2 & 7.0 & 16.0 & 22.4  \\
\hline
$B^{J/\psi}_P(Q^2)$ & 0.25 & 0.23 & 0.21 & 0.19 & 0.18   \\
\hline
$b^{J/\psi}_P(Q^2)$, $GeV^{-2}$ & 0.3 & 0.3 & 0.25 & 0.2 & 0.2   \\
\hline
$B^{J/\psi}_H(Q^2)$ & 0.22 & 0.21 & 0.2 & 0.19 & 0.185   \\
\hline
$b^{J/\psi}_H(Q^2)$, $GeV^{-2}$ & 1.3 & 1.3 & 1.25 & 1.2 & 1.2   \\
\hline
$<|t|>_{W=90\,GeV}$, $GeV^2$ & 0.25 & 0.25 & 0.26 & 0.27 & 0.27   \\
\hline
$A_H$ & \multicolumn{5}{|c|}{2.9}  \\
\hline
$c_H$, $GeV^2$ & \multicolumn{5}{|c|}{0.1}  \\
\hline
\end{tabular}
\end{center}
\caption{Results of fitting 
the data on reaction $\gamma^* + p\to J/\psi + p$.}
\label{tab3}
\end{table}

The slow evolution of $B^{J/\psi}_P(Q^2)$ and $B^{J/\psi}_H(Q^2)$ and 
the decreasing of $b^{J/\psi}_P(Q^2)$ and $b^{J/\psi}_H(Q^2)$ with $Q^2$ 
is also in full accordance with the aforesaid. The hard Pomeron 
trajectory is compared with the soft Pomeron one in Fig. \ref{traV}. 
We point out that besides the fact that the phenomenological trajectory 
$\alpha_H(t)$ has the Kirschner-Lipatov (two-gluon) asymptotics (\ref{hard}), 
its intercept $\alpha_H(0)\approx 1.292$ 
almost coincides with a rough estimation of the lower bound for its 
value obtained in \cite{lipatov}. So, in the diffraction region 
the hard Pomeron contribution 
to the real part of the eikonal is comparable with its contribution to 
the imaginary part in accordance with more general 
theoretical predictions \cite{baranov}. 

\begin{table}[h]
\begin{center}
\begin{tabular}{|l|l|l|l|l|l|l|l|l|l|l|l|}
\hline
$Q^2$, $GeV^2$ & 0.0 & 2.5 & 3.7 & 5.0 & 6.0 & 8.0 & 11.9 
& 13.5 & 19.7 & 32.0 & 41.0  \\
\hline
$B^{\rho^0}_P(Q^2)$ & 3.68 & 3.5 & 3.6 & 3.3 & 3.2 
& 3.0 & 2.6 & 2.5 & 2.0 & 1.7 & 1.5    \\
\hline
$b^{\rho^0}_P(Q^2)$, $GeV^{-2}$ & 0.8 & \multicolumn{10}{|c|}{0.3 (fixed)}   \\
\hline
$B^{\rho^0}_H(Q^2)$ & 0 (fixed) & 0.5 & 0.6 & 0.7 & 0.8 & 0.9 
                                & 1.0 & 1.0 & 1.2 & 1.1 & 1.0    \\
\hline
$b^{\rho^0}_H(Q^2)$, $GeV^{-2}$ &   & \multicolumn{10}{|c|}{1.3 (fixed)}   \\
\hline
\end{tabular}
\end{center}
\caption{Results of fitting 
the data on reaction $\gamma^* + p\to \rho^0 + p$.}
\label{tab4}
\end{table}

At last, after determining the hard Pomeron trajectory 
we can describe the phenomenology of exclusive production of $\rho^0$ 
\cite{rhopho,rhoele}. 
For the case of photoproduction, $Q^2\approx 0$, we can neglect 
the contribution from the hard Pomeron. However, high quality of the 
data on integrated cross sections at nonzero $Q^2$ does not allow us to do 
this for the deeply virtual electroproduction. Besides, determination of 
$t$ slopes of $\bar\beta^{{\rho^0}^*}_H(t,Q^2)$ and  
$\bar\Gamma_P^{{\rho^0}^*}(t,Q^2)$ is quite a hard task 
because of little amount and rather poor-quality data on the 
electroproduction angular distributions. So, for an unambiguous 
description we put the values of these slopes at nonzero $Q^2$ 
equal to those for $J/\psi$-photoproduction, i.e.  
$b^{\rho^0}_P(Q^2)\approx b^{J/\psi}_P(0)=0.3\,GeV^{-2}$ and 
$b^{\rho^0}_H(Q^2)\approx b^{J/\psi}_H(0)=1.3\,GeV^{-2}$. 

\begin{figure}
\epsfxsize=16.7cm\epsfysize=16.7cm\epsffile{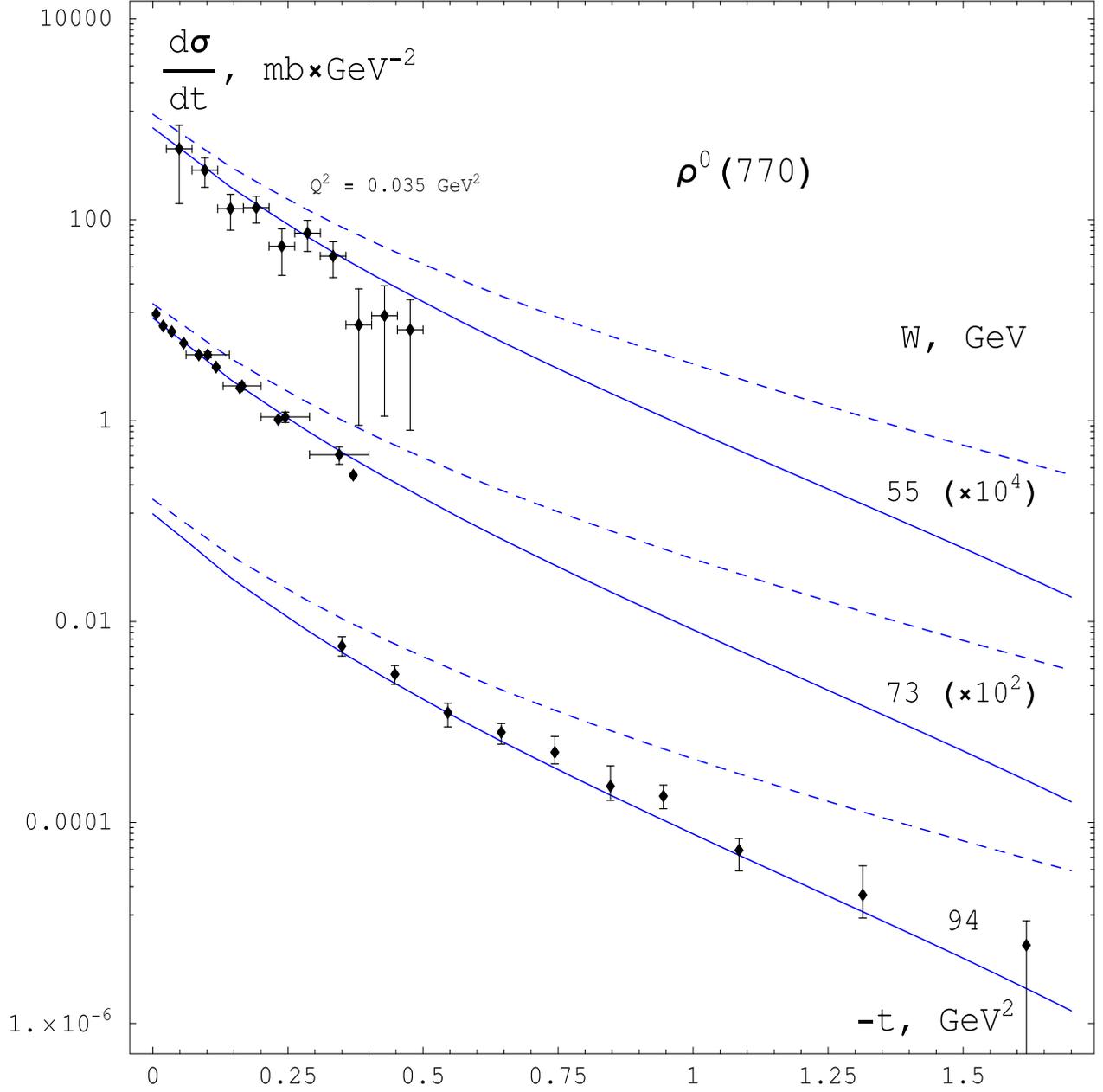}
\caption{Differential cross sections for exclusive 
$\rho^0$-meson photoproduction 
at different values of collision energy (the experimental data and the 
model curve at $W = 55\,GeV$ correspond to the incoming photon 
virtuality $Q^2=0.035\,GeV^2$). Dashed lines correspond to 
cross sections in the Born approximation.}
\label{rho0}
\end{figure}

\begin{figure}
\epsfxsize=16.7cm\epsfysize=16.7cm\epsffile{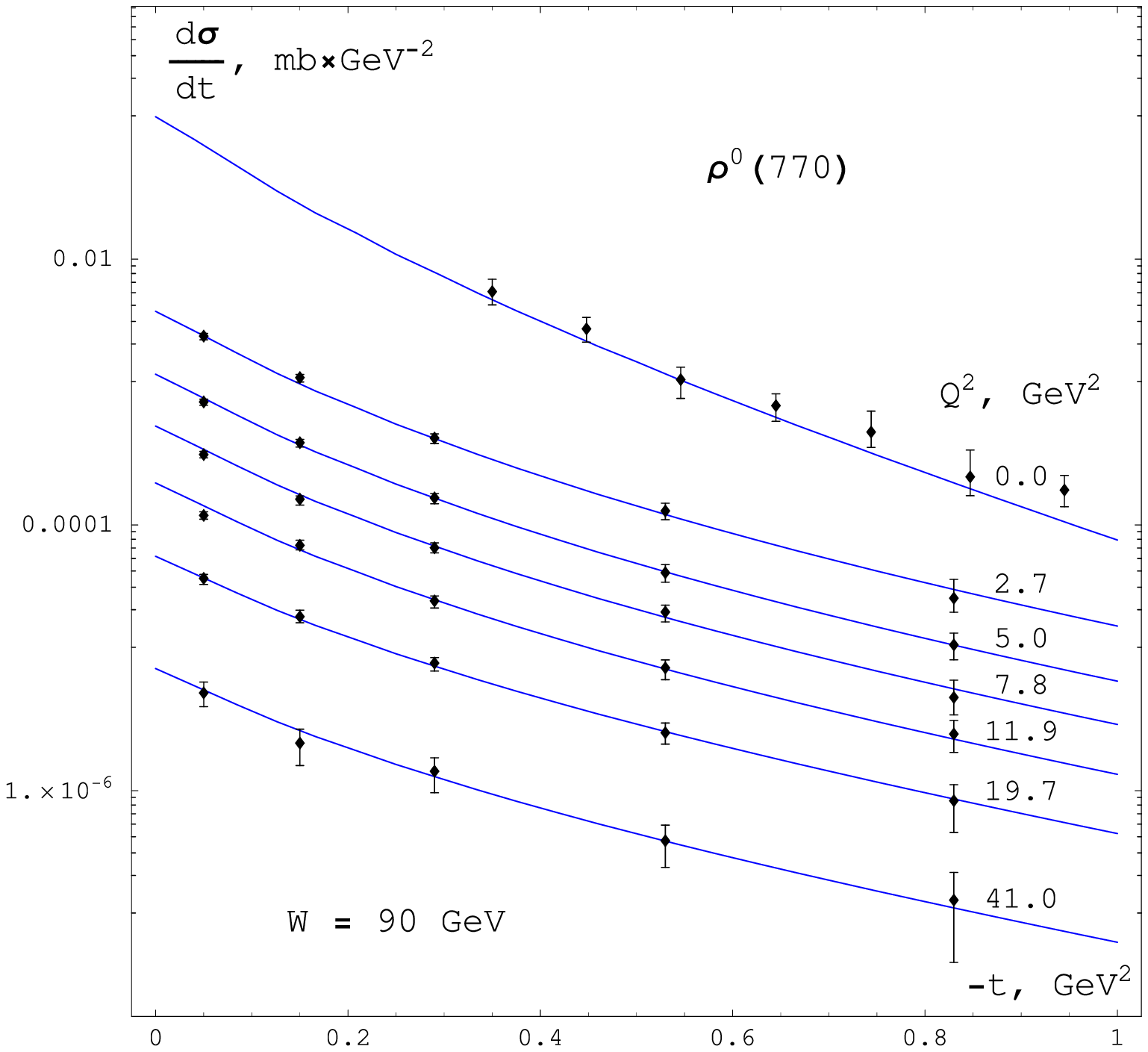}
\caption{Differential cross sections for exclusive 
$\rho^0$-meson electroproduction at collision energy 
$W=90\,GeV$ and different values of the incoming 
photon virtuality.}
\label{diffrho}
\end{figure}

\begin{figure}
\epsfxsize=16.7cm\epsfysize=16.7cm\epsffile{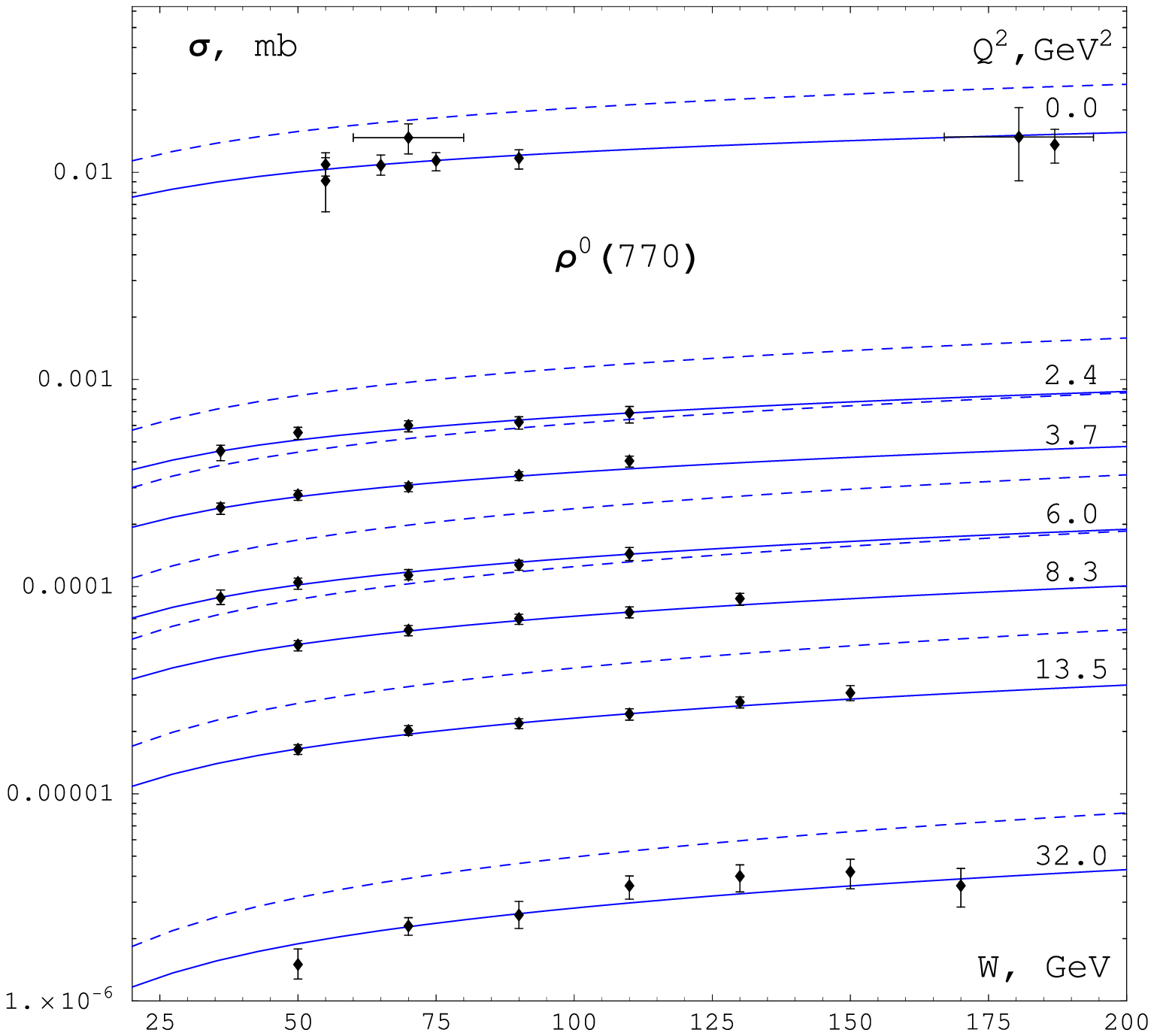}
\caption{Integrated cross sections for exclusive 
$\rho^0$-meson 
electroproduction at different values of the incoming 
photon virtuality as functions of collision energy. 
Dashed lines correspond to 
cross sections in the Born approximation.}
\label{elarho}
\end{figure}

The results of the fitting data are represented in 
Table \ref{tab4} and Figs. \ref{rho0}---\ref{elarho}. 

At the end of this section we would like to emphasize that the main 
success of the used model consists in a satisfactory description of 
electroproduction of $\phi$ meson at $Q^2=5\,GeV^2$ (Fig. \ref{phi5}: 
1 free parameter versus 35 points) 
and photoproduction of $J/\psi$ meson 
(Figs. \ref{jps01}, \ref{jps02}, \ref{jps03}: 6 free parameters 
versus 114 points). 
All other sets of data at fixed $Q^2$ do not, in fact, allow 
unambiguous verification of validity of our Regge 
trajectories because of the insufficient amount of data 
on angular distributions (the data on electroproduction of $\rho^0$ 
and $J/\psi$ do not even allow us to fit parameters $b^V_P(Q^2)$ 
and $b^V_H(Q^2)$ unambiguously). 

\section*{Relation to other models}

Two main features of our approach are the use of an eikonalized (unitarized) 
expression for the scattering amplitude and the exploitation of the 
essentially nonlinear parametrizations of 
Regge trajectories with explicit QCD asymptotic behavior. 

Although there exists a whole sea of papers devoted to the description 
of $W$ and $Q^2$ evolution of integrated cross sections and structure 
functions, only few of them deal with diffractive pattern ($t$ dependence) 
of the scattering amplitude. However, angular distributions contain 
very valuable 
information about the geometry of the interaction region and so they are 
more important quantities than integrated cross sections and structure 
functions. The papers which deal with this subject 
can be conventionally sorted into two groups. 

The first group consists of the purely phenomenological works. 
In \cite{predazzi}, the so-called ``dipole 
Pomeron model'' is used with the soft Pomeron as a double Regge pole 
with a square root threshold singularity. In \cite{fiore}, the model of 
dual amplitudes with Mandelstam analyticity (DAMA) with a square root 
leading singularity is used for the description of the 
$J/\psi$ photoproduction. These two approaches, in fact, do not 
imply the concrete functional dependence of Regge trajectories 
(the authors of these papers use nonlinear parametrizations for 
Regge trajectories) and so are not in contradiction of principle with 
the requirement for Reggeons to have a QCD asymptotic behavior. 
In \cite{donnachie}, the usual Born approximation for the scattering 
amplitude is used with postulated strictly linear Regge trajectories. 
We do not share this point of view since we 
consider the consistency with QCD (which implies nonlinearity) 
more important than quite an arbitrary hypothesis 
about strict linearity. 

The second group consists of two papers \cite{motyka,soyez} 
exploiting the ``colour dipole model'' with 
a main hypothesis that if the incoming photon is of high virtuality 
or the outgoing vector meson is of high mass then the photon fluctuates 
into a quark-antiquark pair which scatters elastically 
off the proton and, at last, recombines into the vector meson. The main 
advantage of this approach is that it allows us to get a concrete 
functional form of $Q^2$ dependence for Reggeon form factors in the 
perturbative region. Hence, the only functional uncertainties in the scattering 
amplitudes is the on-shell Reggeon form factors and Regge trajectories. 
However, these functions must be the same as for the corresponding elastic 
scattering and, so, the concrete applications of the colour dipole approach 
to the vector meson electroproduction must be supplemented 
by the simultaneous description of the high-energy proton-(anti)proton 
diffractive scattering with the same Regge trajectories and form factors of 
the proton. This has not been done in \cite{motyka,soyez}. 
Nevertheless, the colour dipole approach is still the only existing method 
for the quantitative estimation of the $Q^2$ evolution of the 
off shell Reggeon form factors. By no means is it in contradiction 
with the phenomenological approach used in our paper (the colour dipole 
model allows us to determine the $Q^2$ dependence while the Regge-eikonal 
model allows us to determine $W$ and $t$ dependence) and a synthesis 
of these two approaches seems promising for better understanding of 
the diffraction mechanism. 

After obtaining the explicit $Q^2$ dependence of the Reggeon 
form factors in the framework of the colour dipole 
approach the verification of the proposed minimal Regge-eikonal model 
will become possible not only by fitting 
to the HERA data at separate values of $Q^2$ but also by the 
simultaneous fit to the whole 
massive of HERA data on $d\sigma/dt$ and integrated cross sections. 
Moreover, the model will be useful for predictions of cross sections 
of the exclusive vector meson photo- and electroproduction 
in proton-(anti)proton collisions at RHIC, Tevatron, and CERN LHC energies 
\cite{szczurek} (the absorptive corrections due to the soft Pomeron 
exchanges between incoming and between outgoing nucleons may be obtained 
automatically since we know the soft Pomeron trajectory and the 
corresponding form factor of the nucleon).

\section*{Conclusion}

Now we can summarize the results of our analysis. 

The successful simultaneous description of the data on 
elastic proton-(anti)proton scattering \cite{godizov} and the data 
on electroproduction of $\phi$ meson at low values of $Q^2$ 
especially at $Q^2=5\,GeV^2$ in the framework of the one-Pomeron model is 
the evidence that the soft Pomeron absolutely dominates not only 
in the high-energy nucleon-nucleon scattering but also in the low $Q^2$ 
electroproduction of light vector mesons at available 
energies. The results on the description of the high-quality data on 
integrated cross sections of $\rho^0$ electroproduction confirm 
that for electroproduction of light mesons at low $Q^2$ 
(about several $GeV^2$) the contribution of 
the hard Pomeron to the eikonal (Born term) is several times 
less than the soft Pomeron's one. 

The successful description of the $J/\psi$ photoproduction in the 
framework of a two-Pomeron model with the 
hard Pomeron that has the Kirschner-Lipatov asymptotic behavior 
\cite{kirschner} 
in the deeply perturbative domain points to the fact that 
the hard Pomeron has a perturbative nature 
(contrary to the soft Pomeron) and dominates over 
the soft Pomeron at high $Q^2$ and high $W$ in accord with 
expectations of the parton model (the smaller residue slope than 
the soft Pomeron's one results in the domination of the 
hard Pomeron also at larger $t$). 

Among other things, we explicitly demonstrated that neglecting the 
contribution of Regge cuts is not justified {\it a priori} and so in any 
phenomenological model the validity of the Born approximation 
must be grounded by explicit estimation of absorptive corrections.  

Also it was shown that when considering 
the processes of exclusive vector meson production 
there is no need to introduce the ``effective dependence'' of vacuum 
Regge trajectories on the virtuality of the incoming photon 
(this idea is very popular; see, for example, 
\cite{haakman,goloskokov,dosch}). 
In fact, such a dependence contradicts general principles since 
the true Regge trajectories are universal analytical functions 
of one argument and do not depend on the properties of external 
particles (such as virtuality). This hypothesis 
is disguised by the approximation 
$\beta_P(t,Q^2)W^{2\alpha_P(t)}+
\beta_H(t,Q^2)W^{2\alpha_H(t)}\approx
\beta(t,Q^2)W^{2\alpha(t,Q^2)}$ which 
may be valid only in a limited kinematical domain. 

Thus, it was shown that at collision energies higher than $30\,GeV$ 
available data on exclusive diffractive reactions $\gamma^*+p\to V+p$ 
($V = \rho^0,\,\phi,\,J/\psi$), $p+p\to p+p$, 
$\bar p+p\to\bar p+p$ may be described in the framework 
of a simple phenomenological Regge-eikonal model. 
This was achieved by using nonlinear 
Regge trajectories in which their QCD asymptotic 
behavior was taken into account explicitly. 
We would like to point out that linear Regge trajectories not 
only are in contradiction with QCD but even a satisfactory phenomenological 
description of the available data on $t$ dependence 
of exclusive production of vector mesons 
has not been attained yet within approaches using linear 
trajectories (other phenomenological models 
with $Q^2$-independent Regge trajectories and acceptable value of 
$\chi^2/$DOF \cite{predazzi}, \cite{fiore} also use nonlinear 
approximations to them).

\newpage

\end{document}